\documentclass[bibnotes,showpacs,twocolumn,aps,prl,a4paper]{revtex4}

\usepackage{color,epsfig,rotating,graphicx,subfigure}
\usepackage{amsmath,amssymb,amscd}

\usepackage[latin1]{inputenc}
\usepackage[english]{babel}

\usepackage{dsfont}
\usepackage{textcomp}

\renewcommand{\Re}{{\rm Re}}
\renewcommand{\Im}{{\rm Im}}

\newcommand{\ri}{{\rm i}}

\newcommand{\rs}{{\rm s}}
\newcommand{\rp}{{\rm p}}  

\newcommand{\kb}{k_{\rm B}}

\begin{document}

%
%
\title{Towards a black body for near-field thermal radiation} 

\author{S.-A. Biehs and M. Tschikin}

\affiliation{Institut f\"{u}r Physik, Carl von Ossietzky Universit\"{a}t,
D-26111 Oldenburg, Germany.}

\author{P. Ben-Abdallah}

\affiliation{Laboratoire Charles Fabry, Institut d'Optique, CNRS, Universit\'{e} Paris-Sud, Campus
Polytechnique, RD128, 91127 Palaiseau Cedex, France}

\date{\today}


\begin{abstract}
We study the near-field heat exchange between hyperbolic materials and demonstrate 
that these media are able to support broadband frustrated modes which transport 
heat by photon tunnelling with a high efficiency close to the theoretical limit. 
We predict that  hyperbolic materials can be designed to be perfect thermal emitters 
at nanoscale and derive the near-field analog of the black-body limit. 
\end{abstract}

\maketitle
\newpage

%
%

A black body is usually defined by its property of having a maximum absorptivity
and therefore also a maximum emissivity by virtue of Kirchhoff's law~\cite{Planck}. The energy transmission
between two black bodies having different temperatures obey the well-known Stefan-Boltzmann law.
This law sets an upper limit for the power which can be transmitted by real materials, but it is itself
a limit for the far-field only, since it takes only propagating modes into account. 
In terms of the energy transmission between two bodies the black body case corresponds to 
maximum transmission for all allowed frequencies $\omega$ and all wave vectors smaller 
than $\omega/c$, where $c$ is the vacuum light velocity. This means that all the propagating modes 
are perfectly transmitted across the separation gap.  

In the near-field regime, i.e., for distances smaller than the thermal wavelength 
$\lambda_{\rm th} = \hbar c/\kb T$ ($2 \pi \hbar$ is Planck's constant, $\kb$ is Boltzmann's constant, 
and $T$ is the temperature) the radiative heat flux is not due to the propagating modes, but
it is  dominated by evanescent waves~\cite{Polder1971,SurfaceScienceReports,Volokitin2007} and
especially surface polaritons as confirmed by recent 
experiments~\cite{Kittel,Wischnath,HuEtAl2008,NanolettArvind,ShenEtAl2008,NatureEmmanuel,Ottens2011}.
The common paradigm is that the largest heat flux can be achieved when the materials
support surface polaritons which will give a resonant energy transfer restricted to a small 
frequency band around the surface mode resonance frequency~\cite{MuletEtAl2002,MuletAPL,SurfaceScienceReports,Volokitin2007}.
Many researchers have tried to find materials enhancing the nanoscale heat flux due to the contribution
of the coupled surface modes by using layered materials~\cite{Biehs2007,PBA2009}, doped silicon~\cite{Zhang2005,Rousseau}, 
metamaterials~\cite{Joulain2010,Zheng2011,FrancoeurEtAl2011}, phase-change materials~\cite{Zwol2010b} 
and recently graphene~\cite{Svetovoy2011}.

In the present work the aim is twofold: (i) We show, that materials supporting a broad band of evanescent frustrated modes
can outperform the heat flux due to surface modes. This provides new possibilies for designing materials
giving large nanoscale heat fluxes which could be used for thermal management at the 
nanoscale for instance. (ii) We derive a general limit for the heat flux carried by the frustrated modes
and show that it is, in fact, the near-field analog of the usual black body limit. For the evanescent
modes a near field analog of a black body can be defined in the sense that
the energy transmission coefficient must be equal to one for all frequencies 
and all wave vectors larger than $\omega/c$. 

With today's nanofabrication techniques it is possible to manufacture artificial materials 
such as photonic band gap materials and metamaterials which exhibit very unusual 
material properties like negative refraction~\cite{Pendry2004}. Due to such properties
they are considered as good candidates for perfect lensing~\cite{Pendry2000,Larkin2005}, for repulsive 
Casimir forces~\cite{Henkel2005,Leonhardt2007,RosaEtAl2008PRL,RosaEtAl2008} and 
enhanced or tunable radiative heat flux at the nanoscale~\cite{LauEtAl2008,LauEtAl2009,RodriguezEtAl2011,Joulain2010,Zheng2011,FrancoeurEtAl2011} to mention a few. 

There exists a class of uniaxial metamaterials for which the permittivity and permeability
tensor elements are not all of the same sign~\cite{SmithSchurig2003}. In particular, for such 
materials the dispersion relation for the solutions of Helmholtz's equation inside the material is 
not an ellipsoid as for normal uniaxial materials~\cite{Yeh1988} but a 
hyperboloid~\cite{SmithEtAl2004}. For this reason such materials are also called hyperbolic materials. 
These materials have already been considered for super-resolution imaging~\cite{JacobEtAl2006} 
and enhanced thermal conductivity inside the material itself~\cite{NarimanovSmolyaninov2011} for instance. 
Here we focus on the heat flux between two bodies consisting of hyperbolic materials showing that 
these materials can be used to realize a black body at the nanoscale. 

%
%

The nanoscale heat flux $\Phi$ between two semi-infinite materials separated
by a vacuum gap of thickness $d$ having the temperatures $T + \Delta T$ and $T$ can be 
written in terms of the quantum of thermal conductance $\pi^2 \kb^2 T /3 h$~\cite{Pendry1983} 
and the sum over all transversal modes as~\cite{BiehsEtAl2010}
\begin{equation}
  \Phi =  \frac{\pi^2 \kb^2 T}{3 h} \biggl[ \sum_{j = \rs,\rp}
\int \!\! \frac{{\rm d}^2 \kappa}{(2 \pi)^2}  \overline{\mathcal{T}}_j \biggr] \Delta T
\label{Eq:Landauer}
\end{equation}
where $\boldsymbol{\kappa} = (k_x,k_y)^t$ is the lateral wave vector. The mean transmission 
coefficient introduced here is defined as 
\begin{equation}
  \overline{\mathcal{T}}_j = \frac{\int\!\!{\rm d} u\, u^2 e^u (e^u - 1)^{-2} \mathcal{T}_j (u,\kappa) }{\int\!\!{\rm d} u\, u^2  e^u (e^u - 1)^{-2}}
\label{Eq:MeanTransMission}
\end{equation}
where $u = \hbar \omega / \kb T$ is a rescaled frequency or energy and $\mathcal{T}_j(\omega, \kappa)$ 
is the energy transmission coefficient for s- and p-polarized modes in $(\omega,\kappa)$ space. 
For isotropic and anisotropic semi-infinite materials the explicit expression of 
this energy transmission coefficient is known and can be found in Refs.~\cite{Polder1971,BiehsEtAl2011,BiehsEtAl2011b}.

From Eq.~(\ref{Eq:Landauer}) it is obvious that the heat flux $\Phi$ can be increased
on the one hand by increasing the mean transmission coefficient for a transversal
mode with a given $\boldsymbol{\kappa}$ or by increasing the number 
of modes contributing significantly to the heat flux on the other. As discussed 
in Refs.~\cite{Volokitin2007,BiehsEtAl2010}, the number of propagating modes contributing to the heat flux
is limited due to the fact that the dispersion relation of these modes
$\kappa^2 + k_{z0}^2 = \omega^2 / c^2$ restricts for a fixed frequency the modes
to a circle $\kappa \leq \omega/c$. For dielectric materials the frustrated internal reflection modes~\cite{Cravalho}
which can also contribute to the heat flux for distances $d < \lambda_{\rm th}$
the dispersion relation reads $\kappa^2 + k_{z1}^2 = \epsilon \omega^2 / c^2$ so that in this
case the contributing modes are for a fixed frequency restricted to a 
circle $\kappa < \sqrt{\epsilon} \omega/c$.

in presence of surface resonant modes such as surface phonon polaritons

\begin{figure}
  \epsfig{file = 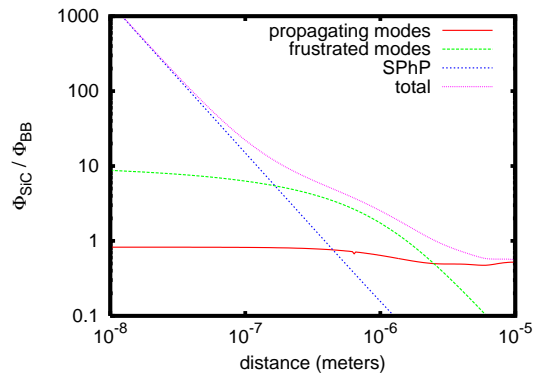, width = 0.4\textwidth}
  \caption{Heat flux $\Phi$ from Eq.~(\ref{Eq:Landauer}) between two semi-infinite SiC bodies over distance
           normalized to the black body value $\Phi_{\rm BB}$ for temperature $T = 300\,{\rm K}$ and $\Delta T = 1\,{\rm K}$. 
           The plot also shows the contribution of the propagating and frustrated modes and the surface 
           phonon polaritons (SPhP).}
  \label{Fig:Modes}
\end{figure}

It is known that in presence of surface resonant modes such as surface phonon polaritons~\cite{KliewerFuchs1974,Shchegrov00} 
in the infrared the heat flux at nanoscale can be increased
by several orders of magnitude~\cite{MuletEtAl2002,MuletAPL} (see Fig.~\ref{Fig:Modes}). In terms of the mean transmission
coefficient $\overline{\mathcal{T}}_j$ it can be shown~\cite{BiehsEtAl2010} that the transmission
of these modes is very small (less than one percent, see Fig.~3 of Ref.~\cite{BiehsEtAl2010}), but the contributing 
modes are for very small distances 
restricted to $\kappa < \log[2/\Im(\epsilon)] /2 d$. First, this means that the number of contributing
modes goes like $1/d^2$ for very small distances so at very small distances the surface phonon polariton contribution to the heat flux
becomes larger than the contribution of the frustrated total internal reflection modes. 
Therefore the surface phonon polaritons provide the dominant heat flux channel at the nanoscale
despite the fact that their transmission coefficient is rather small~\cite{BiehsEtAl2010}. Second, the
upper limit of the wave vectors $\kappa$ for the contributing modes also depends on the 
intrinsic losses~\cite{Pendry1999,BiehsEtAl2010} of the material. 
Note, that there is a fundamental limit restricting $\kappa$ to $\kappa < \pi/a$
where $a$ is for polar materials given by the lattice constant so that there is also 
a definitive limit for the nanoscale heat flux~\cite{Volokitin2004,ZhangJAP}.

\begin{figure}
  \epsfig{file = 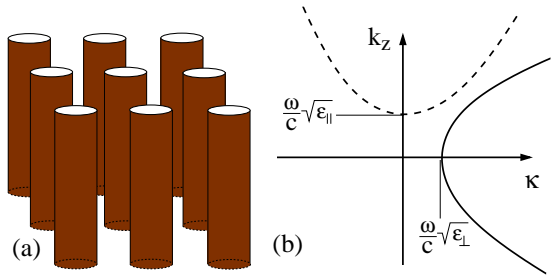, width = 0.4\textwidth}
  \caption{(a) Sketch of a hyperbolic material consisting of a periodical array of nanowires. 
           (b) Hyperbolic dispersion relation in such materials for a fixed frequency. The solid line corresponds 
               to $\epsilon_\parallel <0, \epsilon_\perp >0$ and the dashed line to $\epsilon_\parallel >0, \epsilon_\perp <0$.}
  \label{Fig:Hyperbolic}
\end{figure}

Now we are in a position to discuss the heat flux for hyperbolic 
materials. Such materials are a special class of anisotropic uniaxial materials. 
The dispersion relation for extraordinary waves in uniaxial materials with an optical axis normal to the surface reads~\cite{Yeh1988}
\begin{equation}
  \frac{\kappa^2}{\epsilon_\perp} + \frac{k_{z1}^2}{\epsilon_\parallel} = \frac{\omega^2}{c^2}
\label{Eq:HyperbolicDisp}
\end{equation}
where $\epsilon_\perp$ and $\epsilon_\parallel$ are the permittivity perpendicular and parallel to the surface, respectively, 
having an opposite sign~\cite{SmithSchurig2003}. Hence,
the dispersion relation describes not an ellipse in the $(\kappa,k_z)$ plane but a hyperbolic function as illustrated 
in Fig.~\ref{Fig:Hyperbolic}(b). This means that in principle without losses there is no upper bound for the $\kappa$ which
can fulfill the dispersion relation (\ref{Eq:HyperbolicDisp}). In the following we will
call such modes hyperbolic modes (HM).  If we consider the nanoscale heat flux for such materials 
then the number of contributing modes is only restricted by the intrinsic
cutoff in the energy transmission coefficent which is $\mathcal{T}_j (\omega,\kappa) \propto \exp(- 2 \kappa d)$
for $\kappa \gg \omega/c$ (see Eq.~(\ref{Eq:TransmissionCoeff})). 
It follows that the heat flux due to the HM scales
for small distances like $1/d^2$ as the contribution of the surface phonon polaritons. 
When $\epsilon_\parallel > 0$ and $\epsilon_\perp < 0$ ($\epsilon_\parallel < 0$ and $\epsilon_\perp >0$) 
the HM are propagating inside the hyperbolic material for $\kappa > \omega/c$ ($\kappa > \epsilon_\perp \omega/c $). 
But both kinds of HM are evanescent inside the vacuum region, i.e., for small distances $d \ll \lambda_{\rm th}$ they 
are a special kind of frustrated 
internal reflection modes. As for usual frustrated modes we can expect a large mean transmission coefficient for the 
HM up to $\kappa \approx 1/2d$ if they exist in a broad frequency band. 
Since the mean transmission coefficient for the surface phonon polaritons 
was found to be very small, because these modes are restricted to a small frequency band around the surface mode
frequency,  we can expect to have a heat flux due to the HM larger 
than that of the coupled surface phonon polaritons.

%
%

In order to see whether our expectation is true, we consider a simple example of a uniaxial material which is given
by periodically positioned nanowires~\cite{YaoEtAl2008} of SiC as sketched in Fig.~\ref{Fig:Hyperbolic}(a). 
In the long wave-length limit~\cite{BiehsEtAl2011} we can describe the effective material properties by the 
Maxwell-Garnett expressions~\cite{Saarinen2008}
\begin{align}
  \label{Eq:eps_par}
  \epsilon_\parallel &= \frac{\epsilon(\omega) (1 + f) + (1 - f)}{\epsilon(\omega) (1 - f) + (1 + f)}, \\
  \label{Eq:eps_perp}
  \epsilon_\perp &= (1 - f) + \epsilon(\omega) f,
\end{align}
Here $f$ is the volume filling fraction of the SiC wires and $\epsilon(\omega)$ is the
permittivity of SiC~\cite{Shchegrov00}. 
As pointed out in Ref.~\cite{BiehsEtAl2011} the effective medium theory is limited to distances $d > a/\pi$, where
$a$ is the lattice constant of the nanowires. For smaller distances a nonlocal model for the permittivity would be necessary.

Since the optical axis of the considered material is perpendicular to 
the surface the s- and p-polarized modes decouple and we have
only extraordinary waves for p polarization~\cite{Yeh1988}. Then the energy
transmission coefficient $\mathcal{T}_j$ for two identical half\-spaces which 
enters into Eq.~(\ref{Eq:MeanTransMission}) reads~\cite{BiehsEtAl2011}
\begin{equation}
   \mathcal{T}_{j}(\omega,\kappa; d) =
    \begin{cases}
     (1 - |r_j|^2)^2 / |D_j|^2, & \kappa < \omega/c\\
     4 [\Im(r_j)]^2{\rm e}^{-2 |k_{z0}| d}/|D_j|^2 ,  & \kappa > \omega/c
  \end{cases}
\label{Eq:TransmissionCoeff}
\end{equation}
for $j = \{\rm s, p\}$ where $D_j = 1 - r^2_j {\rm e}^{2 \ri k_{z0} d}$ is the Fabry-P\'{e}rot like denominator,
$k_{z0} = \sqrt{\omega^2/c^2 - \kappa^2}$, and
\begin{equation}
  r_{\rs} (\omega, \kappa)= \frac{k_{z0} - k_\rs}{k_{z0} + k_\rs} \qquad\text{and}\qquad
  r_{\rp} (\omega, \kappa)= \frac{\epsilon_\parallel k_{z0} - k_\rp}{\epsilon_\parallel k_{z0} + k_\rp}
\label{Eq:reflcoeff}
\end{equation}
introducing $k_\rs = \sqrt{\epsilon_{\|}\omega^2/c^2 - \kappa^2}$ and 
$k_\rp = k_{z1} = \sqrt{\epsilon_{\|}\omega^2/c^2 - \frac{\epsilon_{\|}}{\epsilon_{\bot}}\kappa^2}$. From the
expression for the reflection coefficients it becomes obvious that only the extra-ordinary waves
represented by $r_\rp$ are sensitive to $\epsilon_\perp$ and $\epsilon_\parallel$. Hence only the
p-polarized part can support HM because here we are just considering non magnetic materials. 
For magnetic materials also the s-polarized part can support HM.

\begin{figure}
  \epsfig{file = 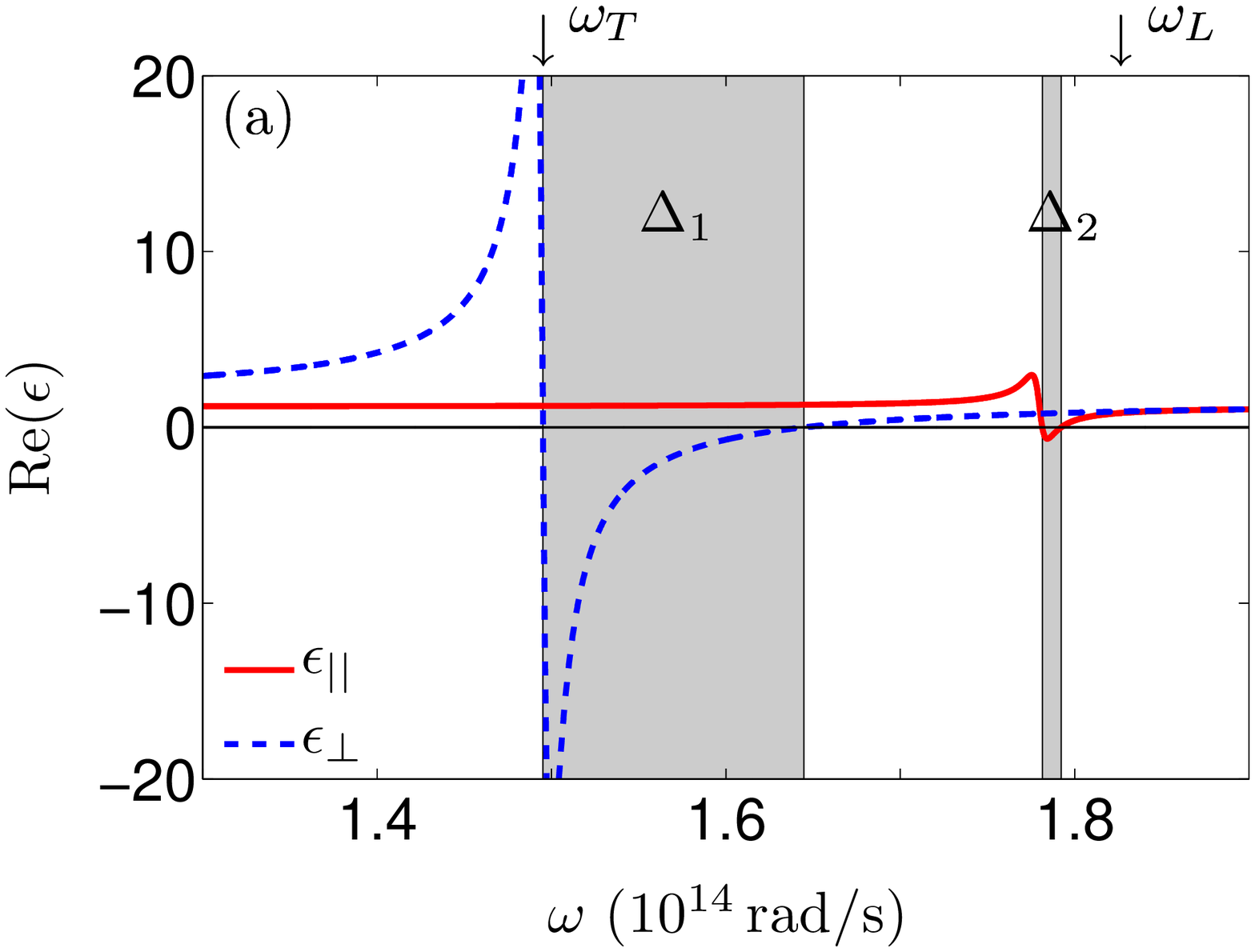, width = 0.4\textwidth}
  \epsfig{file = 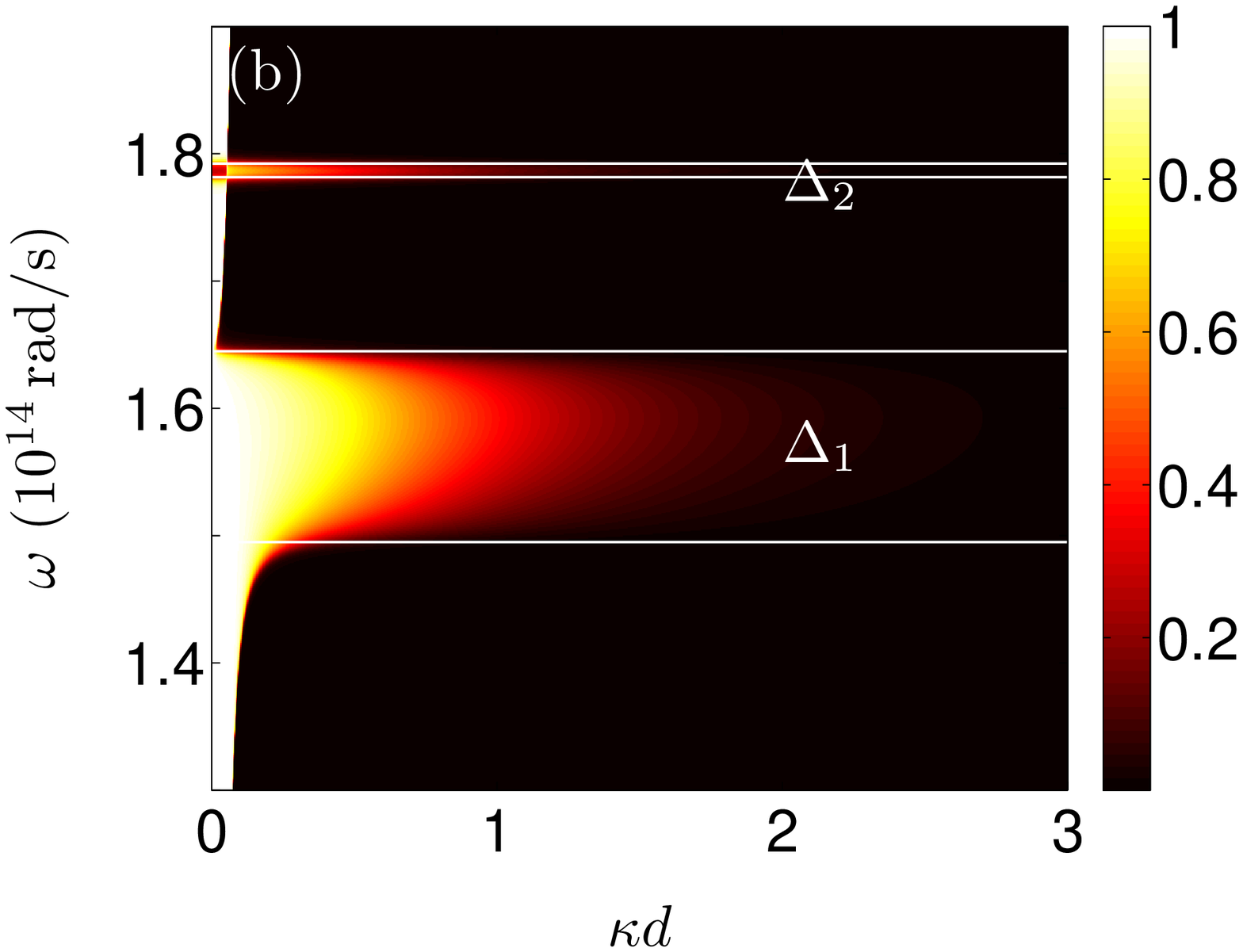, width = 0.39\textwidth}
  \epsfig{file = 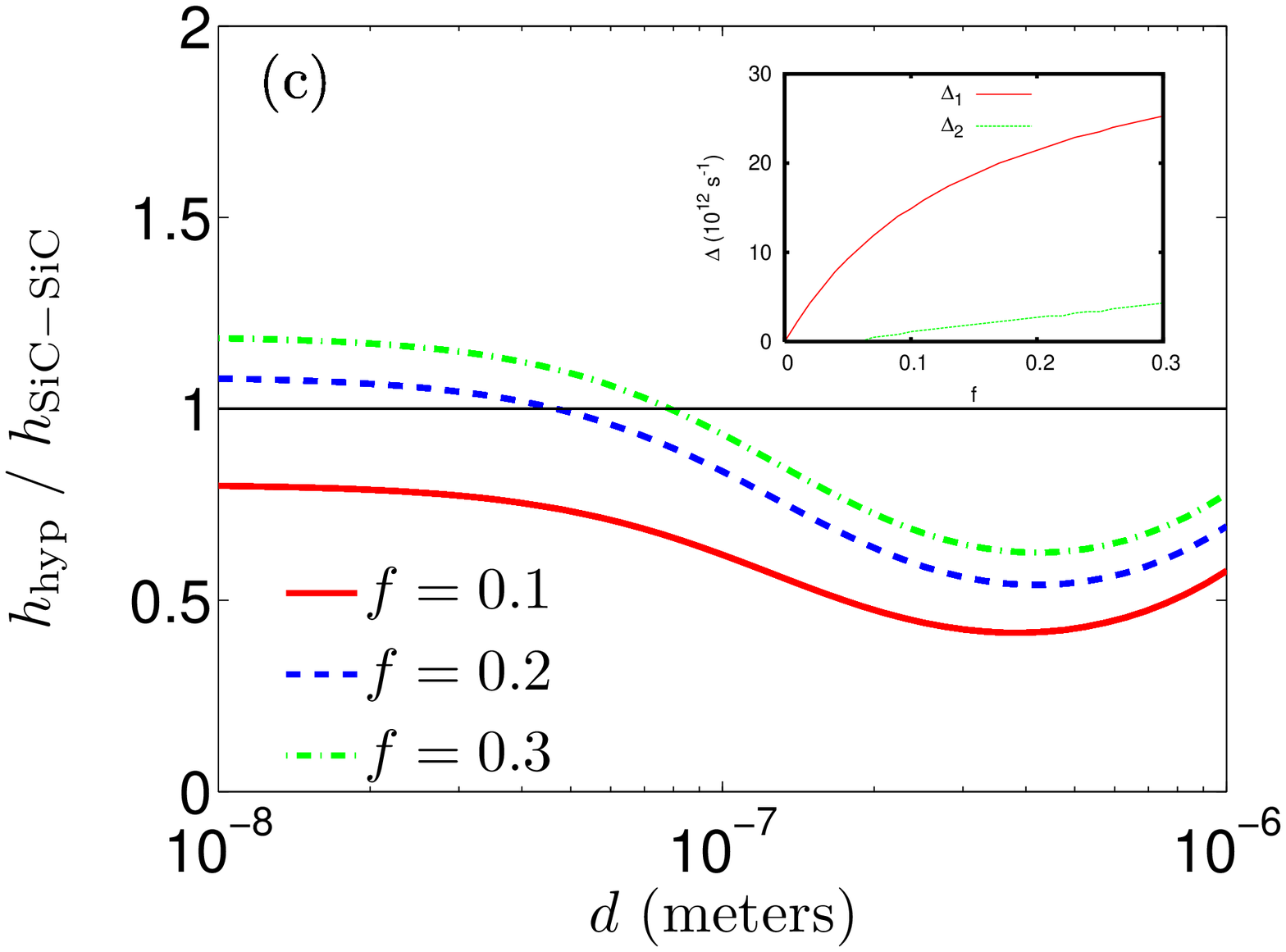, width = 0.39\textwidth}
  \caption{ (a) Real part of $\epsilon_\parallel$ and $\epsilon_\perp$ defined in Eqs.~(\ref{Eq:eps_par}) and (\ref{Eq:eps_perp}) with $f=0.1$. The
                two gray areas mark the frequency region $\Delta _1$ and $\Delta _2$ where both permittivities have different signs, i.e., where the modes
                fullfill a hyperbolic dispersion relation.
            (b) Transmission coefficient $\mathcal{T}_{\rm p}(\omega,\kappa)$ choosing $d = 100\,{\rm nm}$ for the same frequency region and 
                the same volume filling fraction showing high transmission
                for regions $\Delta _1$ and $\Delta _2$.
            (c) Heat transfer coefficient for two hyperbolic materials 
            with $f = 0.1,0.2,0.3$ normalized to the value for two flat SiC 
            half spaces ($T = 300\,{\rm K}$). Inset: $\Delta_1$ and $\Delta_2$ over $f$.
  }
  \label{Fig:EpsTransSiCf01}
\end{figure}

In Fig.~\ref{Fig:EpsTransSiCf01}(a) we plot $\Re(\epsilon_\parallel)$ and $\Re(\epsilon_\perp)$ in a frequency 
range around the transversal and longitudinal phonon frequency  
of SiC ($\omega_{\rm T} = 1.495\cdot 10^{14} {\rm rad} / {\rm s}$  and $\omega_{\rm L} = 1.827\cdot 10^{14} {\rm rad} / {\rm s}$) using $f = 0.1$. The regions $\Delta _1$ and $\Delta _2$ where the permittivities have different signs are highlighted. Within these frequency bands we have HM. In Fig.~\ref{Fig:EpsTransSiCf01}(b) it can be seen that the transmission coefficient $\mathcal{T}_{\rm p} (\omega,\kappa)$ is close to one for these modes up to $\kappa d \approx 1/2$. Furthermore, we observe in Fig.~\ref{Fig:EpsTransSiCf01}(b) that for the chosen parameters we have no coupled surface phonon polaritons in our hyperbolic material so that the heat flux is only due to the usual frustrated modes and the HM. It can be shown that for filling factors $f < 1/3$ there are no surface modes in our structure regardless of the material used for the nanowires.

By increasing the filling factor in our structure the two regions  $\Delta _1$ and $\Delta _2$ supporting
HM get broader (see inset of Fig.~\ref{Fig:EpsTransSiCf01}(c)) 
resulting in larger heat fluxes in the small distance regime around $d = 100\,{\rm nm}$. 
It can be seen in Fig.~\ref{Fig:EpsTransSiCf01}(c) that in this distance regime the heat flux due to
HM can be larger than that by surface phonon polaritons for two SiC media.
This makes clear that the goal for designing a hyperbolic material with elevated heat fluxes is 
to make the bands for HM as broad as possible. This can be done by changing the constituents of the 
hyperbolic material and by considering different geometries like 
for example different nanowire structures or nanolayered materials. 
But there is a general limit for the heat fluxes which can be achieved by such materials.

Now we derive the general theoretical limit for the heat flux due to HM. That means we will
start with the general expression for the heat flux in Eq.~(\ref{Eq:Landauer}) using the reflection coefficients
in Eqs.~(\ref{Eq:reflcoeff}) without assuming a special realization
of the hyperbolic material as the example of the nanowired material. The only assumptions made 
are that we have a uni-axial local medium with its optical axis parallel to the surface normal.  
Before we start, we remind the reader that the usual black body limit can be obtained 
from Eq.~(\ref{Eq:Landauer}) by considering the limits
$\Re(\epsilon_\perp), \Re(\epsilon_\parallel) \rightarrow 1$ and $\Im(\epsilon_\perp), \Im(\epsilon_\parallel) \rightarrow 0$ for all frequencies. Then the $r_{\rm p} \rightarrow 0$ and therefore $\mathcal{T}_{\rm p} \rightarrow 1$
for $\kappa < \omega/c$ and $\mathcal{T}_{\rm p} \rightarrow 0$ for $\kappa > \omega/c$.

With a similar reasoning we will now derive the theoretical limit for near-field radiation by the HM.
For convenience, we consider only the case $\Re(\epsilon_\parallel) > 0$ and $\Re(\epsilon_\perp) < 0$.
Let us first again assume that $\Im(\epsilon_\perp), \Im(\epsilon_\parallel) \rightarrow 0$ for all frequencies
then we can recast the reflection coefficient for p-polarized waves as
\begin{equation}
  r_{\rm p} = \frac{a \ri - 1}{a \ri + 1}
\label{Eq:rp_hm}
\end{equation}
where we have introduced the symbol $a = \sqrt{\epsilon_\parallel |\epsilon_\perp|} \gamma/\gamma_\perp$
with $\gamma^2 = \kappa^2 - \omega^2/c^2$ and $\gamma_\perp^2 = |\epsilon_\perp| \omega^2/c^2 + \kappa^2$.
Since we consider only the frustrated total internal reflection modes ($\kappa > \omega/c$) $a$
is a real number. Hence, as can be expected for total internal reflection modes we have $|r_{\rm p}|^2 = 1$.
That means one can express the reflection coefficient as $r_{\rm p} = \exp(\ri \varphi)$ with 
$\tan(\varphi) = \Im(r_{\rm p})/\Re(r_{\rm p})$. Plugging this reflection coefficient into the transmission
coefficient for the evanescent modes in Eq.~(\ref{Eq:TransmissionCoeff}) we find that $\mathcal{T}_{\rm p} = 1$ 
for $|k_{z0}| d \ll 1$. On the other hand for $|k_{z0}| d \gg 1$  we find that $\mathcal{T}_{\rm p}$ 
has a maximum for $r_{\rm p} = \pm \ri$. It follows from Eq.~(\ref{Eq:rp_hm}) that the condition for having 
perfect transmission is $a^2 = 1$ which translates in the quasi-static 
regime ($\kappa \gg \omega/c$) into $\epsilon_\parallel |\epsilon_\perp| = 1$.

Finally, we can derive the maximum heat flux due to the frustrated HM by plugging $r_{\rm p} = \pm \ri$ into
the expression for the transmission coefficient in Eq.~(\ref{Eq:TransmissionCoeff}) and evaluating 
Eq.~(\ref{Eq:Landauer}) for the heat flux. Then, we find a maximum heat transfer coefficient ($\Phi = h \Delta T$) of
\begin{equation}
   h_{\rm max, p} = \frac{1}{\pi d^2}\biggl( \frac{\pi^2 \kb^2 T}{3 h}\biggr)  \frac{\ln(2)}{2}.
\label{Eq:hmax}
\end{equation}
This expression constitutes the black body limit for the near-field radiation by HM, since it is due to 
perfect transmission ($\mathcal{T}_{\rm p} = 1$ for $\omega/c < \kappa \ll 1/(2d)$) by photon tunneling. Note, that
this limit is by a factor $\ln(2)/2$ smaller than the upper limit found 
in Ref.~\cite{JoulainPBA2010}, which was derived by making an assumption on the cutoff in the $\kappa$ 
integration and is therefore ambiguous.

Let us now compare the upper limit in Eq.~(\ref{Eq:hmax}) with known results. For $T = 300\,{\rm K}$ the
heat transfer coefficient for a black body is about $6.1\,{\rm W}{\rm m}^{-2}{\rm K}^{-1}$.
According to Eq.~(\ref{Eq:hmax}), its near-field counterpart ($h_{\rm max, p} \approx 3.3\cdot10^5\,{\rm W} {\rm m}^{-2}{\rm K}^{-1}$)
is about $5.4\cdot10^4$ times larger at $d = 10\,{\rm nm}$.
For the same distance the heat transfer coefficient between two gold plates is 
about $1157\,{\rm W}{\rm m}^{-2}{\rm K}^{-1}$ and between two SiC plates 
one gets $9200\,{\rm W}{\rm m}^{-2}{\rm K}^{-1}$. Hence $h_{\rm max, p}$ is by a factor of 
about $285$ or $36$ larger, resp. Finally, we compare the limit for the heat transfer coefficient due to 
photon tunneling of HM from Eq.~(\ref{Eq:hmax}) with a numerically found upper limit for Drude materials 
for the heat transfer coefficient due to the coupling of surface plasmon polaritons 
(see parameter set A in Ref.~\cite{WangEtAl}) which is about $3\cdot10^4$ times the black body value 
at $d = 10\,{\rm nm}$. That means the limit in Eq.~(\ref{Eq:hmax}) is by 
a factor $5/3$ larger than the value one can get by tuning surface mode resonances. Hence, the design 
of hyperbolic materials constitutes a new way for an efficient enhancement of the nanoscale heat flux.

%
%

In conclusion, we have studied the near-field heat transfer between hyperbolic materials 
and we have shown with a simple example that the broadband spectrum of frustrated modes 
supported by these media allows us to get heat fluxes which are larger than the heat flux
due to narrow-band coupled surface polariton modes. We have derived a theoretical limit of
the heat flux due to hyperbolic modes and demonstrated that it constitutes the near-field analog of
the black-body limit. This opens up new possibilities to achieve large heat fluxes at the nanoscale by the design
of hyperbolic materials.  

%
%

\begin{acknowledgments}
\end{acknowledgments}

%
%

\appendix

%
%


\begin{thebibliography}{99}
  \bibitem{Planck} M. Planck, {\em The theory of heat radiation}, (Forgotten Books,Leipzig,2010). \
  \bibitem{Polder1971} D. Polder and M. Van Hove, Phys. Rev. B \textbf{4} 3303 (1971).
  \bibitem{SurfaceScienceReports} K. Joulain, J.-P. Mulet, F. Marquier, R. Carminati, and J.-J. Greffet, Surf. Sci. Rep. \textbf{57}, 59-112 (2005).
  \bibitem{Volokitin2007} A. I. Volokitin and B. N. J. Persson, Rev. Mod. Phys. \textbf{79}, 1291 (2007).
  \bibitem{Kittel} A. Kittel, W. M\"uller-Hirsch, J. Parisi, S.A. Biehs, D. Reddig, and M. Holthaus, Phys. Rev. Lett. \textbf{95}, 224301 (2005).
  \bibitem{Wischnath} U. F. Wischnath, J. Welker, M. Munzel and A. Kittel, Rev. Sci. Instrum. \textbf{79}, 073708 (2008).
  \bibitem{HuEtAl2008} L. Hu, A. Narayanaswamy, X. Chen, and G. Chen, \apl {\bf 92}, 133106 (2008).
  \bibitem{NanolettArvind} A. Narayanaswamy, S. Shen, and G. Chen, Phys. Rev. B \textbf{78}, 115303 (2008).
  \bibitem{ShenEtAl2008} S. Shen, A. Narayanaswamy, and G. Chen, Nano Lett. {\bf 9}, 2909 (2009).
  \bibitem{NatureEmmanuel} E. Rousseau, A. Siria, G. Jourdan, S. Volz, F. Comin, J. Chevrier and J.-J. Greffet,  Nature Photonics {\bf 3}, 514 (2009).
  \bibitem{Ottens2011} R. S. Ottens, V. Quetschke, S. Wise, A. A. Alemi, R. Lundock, G. Mueller, D. H. Reitze, D. B. Tanner, B. F. Whiting, Phys. Rev. Lett. {\bf 107}, 014301 (2011).
  \bibitem{MuletEtAl2002} J.-P. Mulet, K. Joulain, R. Carminati, and J.-J. Greffet, Microscale Thermophysical Engineering {\bf 6}, 209 (2002).
  \bibitem{MuletAPL}J.-P. Mulet, K. Joulain, R. Carminati, and J.-J. Greffet, Appl. Phys. Lett. \textbf{78}, 2931 (2001).
  \bibitem{Biehs2007} S.-A. Biehs, Eur. Phys. J. B {\bf 58}, 423-431 (2007).
  \bibitem{PBA2009} P. Ben-Abdallah, Karl Joulain, J. Drevillon, and G. Domingues, J. Appl. Phys. {\bf
106}, 044306 (2009).
  \bibitem{Zhang2005} C.J. Fu and Z. M. Zhang, Int. J. Heat Mass Transfer {\bf 49}, 1703 (2006).
  \bibitem{Rousseau} E. Rousseau, M. Laroche, and J.-J. Greffet, Appl. Phys. Lett. {\bf 95}, 231913 (2009).
  \bibitem{Joulain2010} K. Joulain, J. Drevillon, and P. Ben-Abdallah, Phys. Rev. B {\bf 81}, 165119 (2010).
  \bibitem{Zheng2011} Z. Zheng and Y. Xuan, Chin. Sci. Bull. {\bf 56}, 2312 (2011).
  \bibitem{FrancoeurEtAl2011} M. Francoeur, S. Basu, and S. J. Petersen, Opt. Expr. {\bf 19}, 18774 (2011).
  \bibitem{Zwol2010b} P. J. van Zwol, K. Joulain, P. Ben-Abdallah, and J. Chevrier, Phys. Rev. B {\bf 84}, 161413 (2011). 
  \bibitem{Svetovoy2011} V. B. Svetovoy, P. J. van Zwol, J. Chevrier, arXiv:1201.1824v1.
  \bibitem{Pendry2004} J. B. Pendry, Contemp. Phys. {\bf 45}, 191 (2004).
  \bibitem{Pendry2000} J. B. Pendry, Phys. Rev. Lett. {\bf 85}, 3966 (2000).
  \bibitem{Larkin2005} I. A. Larkin and M. I. Stockman, Nano Lett. {\bf 5}, 339 (2005).
  \bibitem{Henkel2005} C. Henkel and K. Joulain, Eur. Phys. Lett. {\bf 72}, 929 (2005).
  \bibitem{Leonhardt2007} U. Leonhardt and T. G. Philbin, New. J. Phys. {\bf 9}, 254 (2007).
  \bibitem{RosaEtAl2008PRL} F. S. S. Rosa, D. A. R. Dalvit, and P. W. Milonni, Phys. Rev. Lett. {\bf 100}, 183602 (2008).  
  \bibitem{RosaEtAl2008} F. S. S. Rosa, D. A. R. Dalvit, and P. W. Milonni, Phys. Rev. A {\bf 78}, 032117 (2008).
  \bibitem{LauEtAl2008} W. T. Lau, J.-T. Shen, G. Veronis, S. Fan, and P. V. Braun, Appl. Phys. Lett. {\bf 92}, 103106 (2008).
  \bibitem{LauEtAl2009} W. T. Lau, J.-T. Shen, S. Fan, Phys. Rev. B {\bf 80}, 155135 (2009).
  \bibitem{RodriguezEtAl2011} A. W. Rodriguez, O. Ilic, P. Bermel, I. Celanovic, J. D. Joannopoulos, M. Soljacic, and S. G. Johnson, Phys. Rev. Lett. {\bf 107}, 114302 (2011).
  \bibitem{SmithSchurig2003} D. R. Smith and D. Schurig, Phys. Rev. Lett. {\bf 90}, 077405 (2003).
  \bibitem{Yeh1988} P. Yeh, {\em Optical waves in layered media}, (Wiley, New York, 1988).
  \bibitem{SmithEtAl2004} D. R. Smith, P. Kolinko, and D. Schurig, J. Opt. Soc. B {\bf 21}, 1032 (2004).
  \bibitem{JacobEtAl2006} Z. Jacob, L. V. Alekseyev, and E. Narimanov, Opt. Exp. {\bf 14}, 8247 (2006).
  \bibitem{NarimanovSmolyaninov2011} E. E. Narimanov and I. I. Smolyaninov, arXiv:1109.5444v1.
  \bibitem{Pendry1983} J. B. Pendry, J. Phys. A: Math. Gen. \textbf{16}, 2161 (1983).
  \bibitem{BiehsEtAl2010} S.-A. Biehs, E. Rousseau, and J.-J. Greffet, Phys. Rev. Lett. \textbf{105}, 234301 (2010).
  \bibitem{BiehsEtAl2011} S.-A. Biehs, P. Ben-Abdallah, F. S. S. Rosa, K. Joulain, and J.-J. Greffet, Opt. Expr. {\bf 19}, A1088-A1103 (2011).
  \bibitem{BiehsEtAl2011b} S.-A. Biehs, P. Ben-Abdallah, and F. S. S. Rosa, Appl. Phys. Lett. {\bf 98}, 243102 (2011).
  \bibitem{Cravalho} E. G. Cravalho, C. L. Tien, and R. P. Caren, J. Heat Transfer \textbf{89}, 351 (1967).
  \bibitem{KliewerFuchs1974} K. L. Kliewer and R. Fuchs, Adv. Chem. Phys. {\bf 27}, 355 (1974).
  \bibitem{Shchegrov00} A. V. Shchegrov, K. Joulain, R. Carminati, and J.-J. Greffet, \prl {\bf 85}, 1548 (2000).
  \bibitem{Pendry1999} J. B. Pendry (1999), J. Phys.: Conden. Mat. {\bf 11}, 6621 (1999).
  \bibitem{Volokitin2004} A. I. Volokitin and B. N. J. Persson, Phys. Rev. B \textbf{69}, 045417 (2004).
  \bibitem{ZhangJAP} S. Basu and Z. Zhang, J. Appl. Phys. \textbf{105}, 093535 (2009).
  \bibitem{YaoEtAl2008} see for instance: J. Yao, Z. Liu, Y. Liu, Y. Wang, C. Sun,G. Bartal, A. M. Stacy, and X. Zhang {\bf 321}, 930 (2008).
  \bibitem{Saarinen2008} J. J. Saarinen, S. M. Weiss, P. M. Fauchet, and J. E. Sipe, J. Appl. Phys. {\bf 104}, 013103 (2008).
  \bibitem{JoulainPBA2010} P. Ben-Abdallah and K. Joulain, Phys. Rev. B \textbf{82}, 121419(R) (2010).
  \bibitem{WangEtAl} X. J. Wang, S. Basu, and Z. M. Zhang, J. Phys. D: Appl. Phys. {\bf 42}, 245403 (2009).
\end{thebibliography}
\end{document}